\documentclass[sigconf]{acmart}

\AtBeginDocument{%
  \providecommand\BibTeX{{%
    \normalfont B\kern-0.5em{\scshape i\kern-0.25em b}\kern-0.8em\TeX}}}

\usepackage{url}
\usepackage{amsmath,amsfonts} 
\usepackage{float}
\usepackage{graphicx}
\usepackage{algorithm}
\usepackage{algpseudocode}
\usepackage{caption}
\usepackage{subcaption}
\usepackage{multirow}
\usepackage{adjustbox}
\usepackage{makecell}

\usepackage{enumitem}

\begin{document}

\title{A Testbed To Study Adversarial Cyber-Attack Strategies in Enterprise Networks}


\author{Ayush Kumar}
\email{ayush_kumar@sutd.edu.sg}
\orcid{0000-0002-5174-9906}
\affiliation{%
  \institution{Singapore University of Technology and Design}
}

%
\author{David K. Yau}
\email{david_yau@sutd.edu.sg}
\affiliation{%
  \institution{Singapore University of Technology and Design}
}


\begin{abstract}
In this work, we propose a testbed environment to capture the attack strategies of an adversary carrying out a cyber-attack on an enterprise network. The testbed contains nodes with known security vulnerabilities which can be exploited by hackers. Participants can be invited to play the role of a hacker (e.g., black-hat, hacktivist) and attack the testbed. The testbed is designed such that there are multiple attack pathways available to hackers. We describe the working of the testbed components and discuss its implementation on a VMware ESXi server. Finally, we subject our testbed implementation to a few well-known cyber-attack strategies, collect data during the process and present our analysis of the data.
\end{abstract}



\maketitle

\section{Introduction}
Cyber-attacks are a major factor affecting the operations of enterprises, government organizations, critical infrastructures, SMEs, etc. Not only do they threaten to disrupt those operations, the attacks are a leading cause of financial losses to the affected organizations\cite{varonis-cybersec-stats, purplesec-cybersec-stats}. A number of cyber-attacks are targeted at enterprise networks belonging to SME (Small and Medium Enterprises) as well as large enterprises, e.g., to steal valuable business data. When an adversary targets such a network, their attack strategy leaves a footprint across the network and the devices connected to it. By attack strategy, we refer to actions such as how an adversary gains access to a target network, how it moves across the network, how it looks for vulnerable machines connected to the network and exploits them, etc. If data is collected from the target network when under attack by an adversary, it can be analyzed for patterns. Further, those patterns can be used to detect future cyber-attacks. Data collected from our testbed can also be used to validate frameworks modelling the motivations, cognitive antecedents and dynamic decision making processes of hackers in the lead up to as well as during cyber-attacks which can be helpful in predicting them. As we can not allow an adversary to attack a real enterprise network, it makes sense to build a testbed environment that emulates an enterprise network and let adversaries attack the testbed instead.

Hence, in this work, we present a controlled testbed environment to emulate enterprise networks and capture adversarial attack strategies. The testbed consists of nodes configured to serve functions similar to the machines that can be found in a real-world enterprise network. Some of the nodes are configured with known security vulnerabilities. Adversaries can be allowed to attack the testbed network by exploiting the vulnerabilities and the resulting system and network-level data can be collected using an in-built data logging infrastructure. This testbed is targeted at cyber security researchers who can easily and quickly bootstrap experiments. It also offers flexibility in terms of changing the testbed network topology, modifying existing nodes and their functionalities or increasing the number of nodes when required.


\section{Related Work}
\label{literature}
There are only a handful of studies on building enterprise network testbeds in existing literature. Few works\cite{vnetwork, scan-worm-enterprise} have proposed testbeds using a node virtualization approach to emulate real-world enterprise networks which can be used for analysing worm/malware propagation. 
These testbeds are supposed to be used to evaluate/validate detection and defense techniques against malware/worm propagation. However, they are more focused on a specific cyber-attack (malware, worms) and emulating the topology and number of nodes in real-world enterprise networks. In comparison, our focus is on capturing the architecture of real-world enterprise networks. Moreover, our proposed testbed is aimed at studying the attack strategies of cyber adversaries and using the patterns extracted from them to design cyber-attack detection methods in enterprise networks.

In \cite{rel-enterprise-netw}, the authors have presented a testbed to evaluate their proposed fault-localization system, Sherlock, meant to be deployed in large enterprise networks. The testbed consists of two LANs connected by a router, with each LAN containing web servers, SQL backend server, DNS servers, authentication servers and clients. Their testbed is closest to ours in terms of architecture, though its purpose is entirely different.

\section{Testbed Overview}
\label{testbed-overview}
In this section, we begin by giving an overview of the architecture and configuration of our proposed testbed followed by listing down the security vulnerabilities used to infect some of the testbed nodes.

\subsection{Architecture and Configuration}
\begin{figure}[h]
\centering
\includegraphics[scale=0.3]{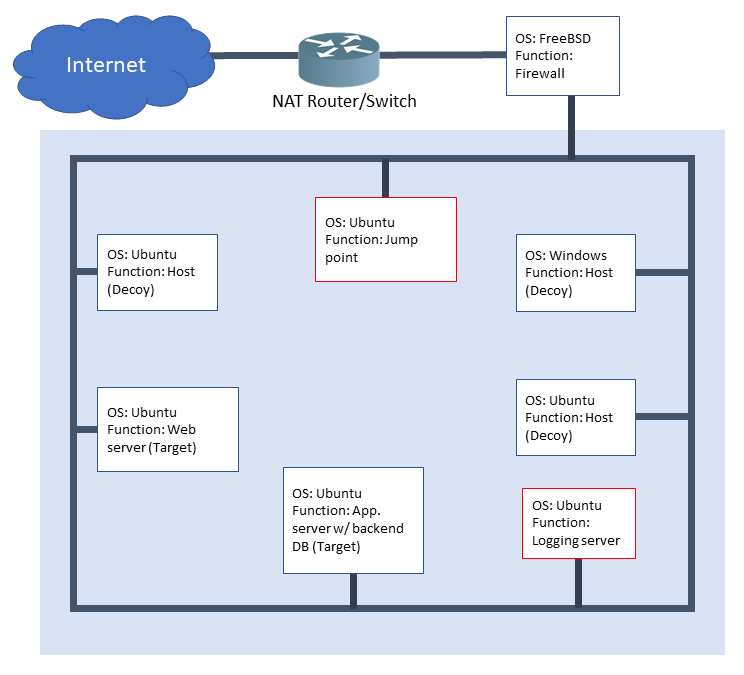}
\caption{Testbed Architecture (Small and Medium Enterprise)}
\label{testbed-arch1}
\end{figure}

\begin{figure}[h]
\centering
\includegraphics[scale=0.3]{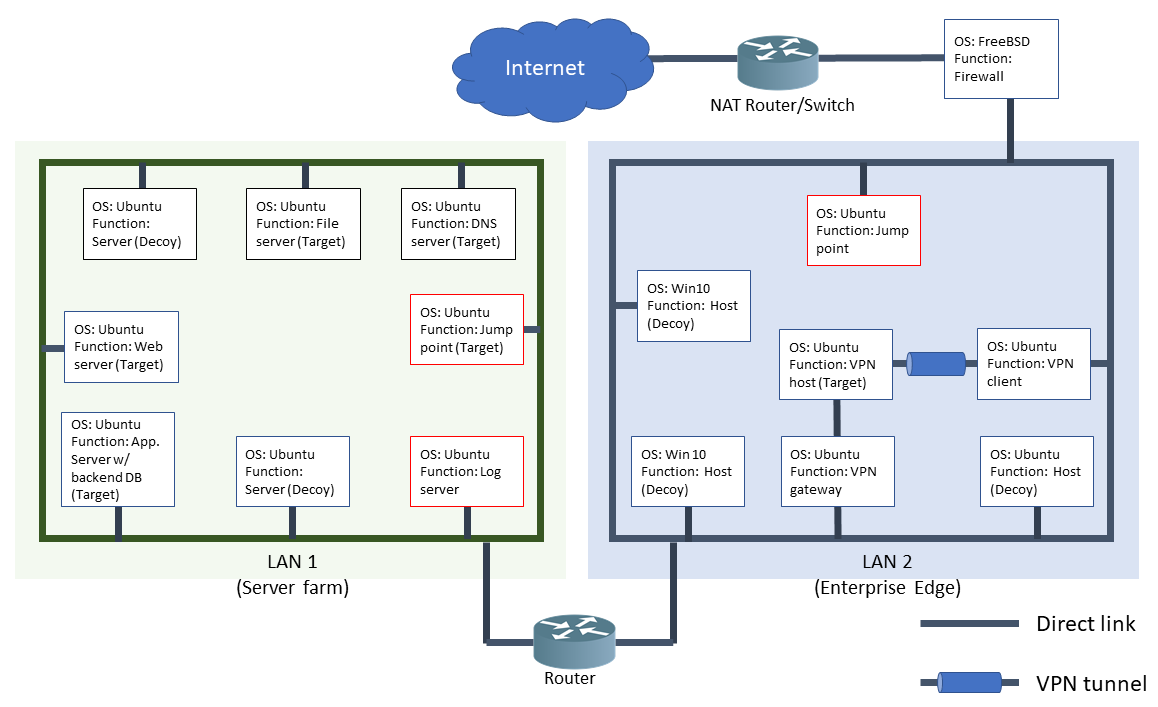}
\caption{Testbed Architecture (Large Enterprise)}
\label{testbed-arch2}
\end{figure}

Large enterprises and SMEs (Small and Medium Enterprises) have different network architectures. Therefore, we have proposed two sets of architectures for our testbed representing large enterprises and SMEs' networks as shown in Fig. \ref{testbed-arch1} and Fig. \ref{testbed-arch2}. The testbed machines and their functions are listed below:
\begin{itemize}
\item \textbf{Jump point/host}: This is the machine through which a hacker gains entry into our testbed network.
\item \textbf{Web server}: This machine acts as a web server hosting a fictitious company's web page.
\item \textbf{Application server}: This machine is configured as a server hosting a web application with a MySQL database at its back-end. Through the web application, employees can login with their credentials to retrieve and update their employment-related details saved with the company. The database also stores the VPN (Virtual Private Network) credentials of the employees which are used to connect to hosts connected to a remote company site.
\item \textbf{VPN client}: This machine acts as a VPN client which is used to connect to company's VPN network. 
\item \textbf{VPN server}: This machine runs a VPN server, waiting for connection requests from VPN clients. The TLS/SSL VPN supports packet encryption, certificate-based server validation, client authentication and multiple clients. 
\item \textbf{VPN host}: This machine acts as a host accessible from the VPN network only. It stores classified files and documents on the company's products.
\item \textbf{DNS server}: This machine runs a local DNS server, which maps internal company domains to IP addresses as well as handles external DNS queries.
\item \textbf{File server}: This machine runs a local file server storing important files which can be accessed by the company's employees.
\item \textbf{Logging server}: This machine is the centralized point where system logs and packet traces from all the machines connected to the testbed network are collected. It has two modules:
	\begin{itemize}
	\item A packet traffic capture utility (e.g., Wireshark, Tcpdump) runs on the logging server in promiscuous mode so that it listens to all the packet traffic on the testbed network being sent or received by the other machines.
	\item The system log (\textit{syslog}) files from the all the machines are piped to the logging server using \textit{rsyslog} utility.
	\end{itemize}
\item \textbf{Decoy hosts}: These machines are meant to confuse the hackers and delay their attack on target machines. 
\end{itemize}

\subsection{Security Vulnerabilities}
We have included several real-world security vulnerabilities in the testbed nodes which can be exploited by hackers to accomplish their attacks on our testbed. The security vulnerabilities are listed below:
\begin{itemize}
\item The web server is configured with a weak login password which hackers should be able to obtain using password brute forcing.
\item The MySQL database at the backend of the application server is vulnerable to SQL injection. This means that using specially crafted inputs to the web application, hackers can retrieve all or some of the employees' sensitive information stored in the database.
\item The VPN server uses only password-based authentication to authenticate clients which can result in hackers obtaining access to the VPN network using stolen employee VPN credentials.
\item The DNS server is vulnerable to cache poisoning in which a hacker can spoof responses from other DNS servers to redirect employees to malicious domains.  
\item The file server is vulnerable to remote command execution, i.e., a hacker can obtain a reverse-shell to the file server. 
\end{itemize}

\section{Hacking the Testbed}
In this section, we present some of the possible pathways which can be taken by hackers in our testbed. 

\subsection{Attack pathways}
\label{pathways}
In our testbed, hackers have been provided choices in terms of nodes with different functionalities and their associated security vulnerabilities. It is up to the hacker to figure how to reach the nodes, decide whether to attack them and the kind of attack to carry out. We now outline some of the possible attack pathways that can be followed in our testbed by three types of hackers: hacktivists\cite{caldwell-2015}, petty thieves\cite{peters-2015} and black-hats\cite{kaspersky-2018}.

\subsubsection{Hacktivist pathways}
\begin{figure}[h]
\centering
\includegraphics[scale=0.25]{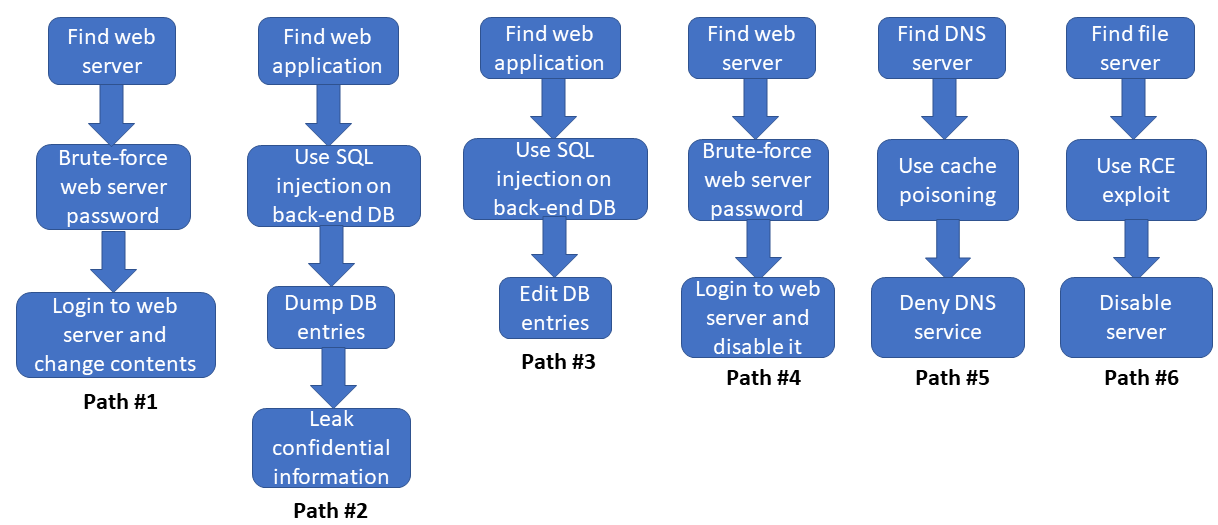}
\caption{Possible testbed pathways for a Hacktivist}
\label{hacktivist-pathways}
\end{figure}

Hacktivists are highly skilled and generally tend to disrupt/damage systems or leak confidential information. Therefore, they can target both SME and large enterprise networks, though in the latter case they may have to use malware to spread and infect machines, use privilege escalation techniques and move across the company network (LAN2 $\rightarrow$ LAN1). Hacktivists may attempt to deface the website hosted on web server by changing its contents, or leak private employee details online, or change the contents of the database at the back-end of the application server, or disable the web server itself, or deny DNS service, or disable the file server. The pathways corresponding to hacktivist are shown in Fig. \ref{hacktivist-pathways}. 

\subsubsection{Petty Thief pathways}
\begin{figure}[h]
    \centering
    \begin{subfigure}[b]{0.4\textwidth}
        \centering
        \includegraphics[width=\textwidth]{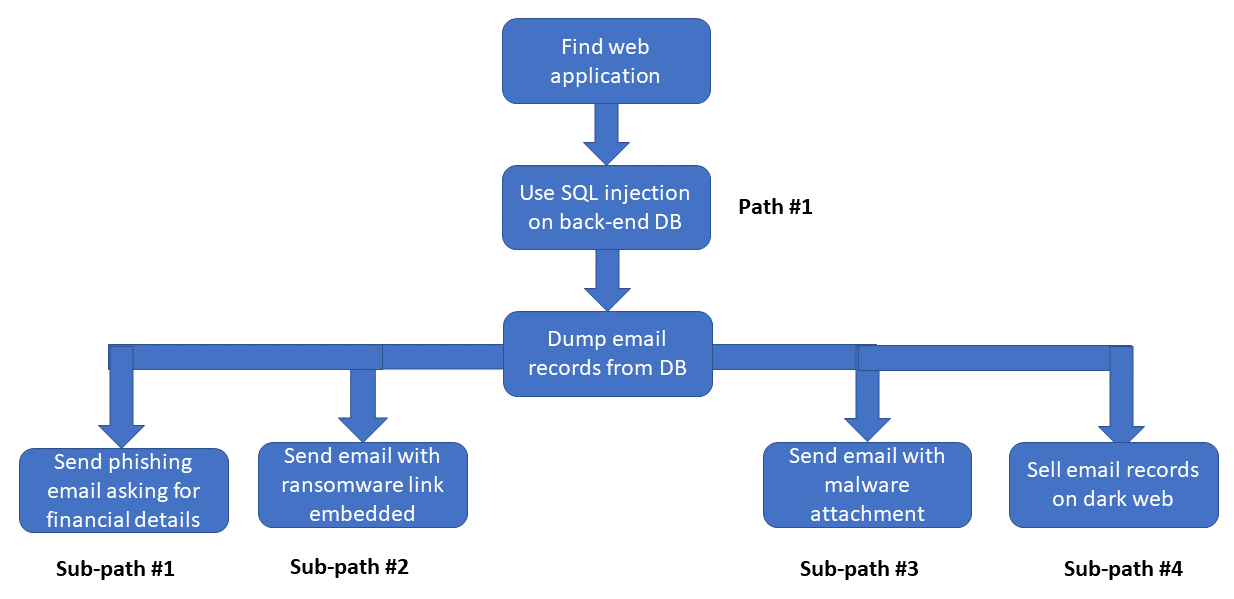}
        \caption{}
        \label{}
    \end{subfigure} 
    \par\medskip
    \begin{subfigure}[b]{0.4\textwidth}
        \centering
        \includegraphics[width=\textwidth]{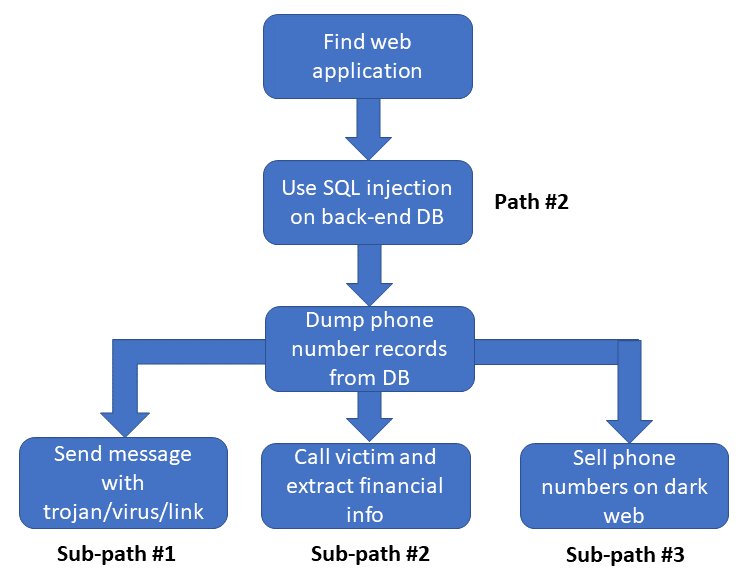}
        \caption{}
        \label{}
    \end{subfigure}
    \caption{Possible testbed pathways for a Petty Thief}
    \label{petty-thief-pathways}
\end{figure}

Since petty thieves are mostly financially motivated and low to medium-skilled, they would target an SME network rather than a large enterprise network. They may attempt to obtain e-mail and phone records of the employees from the database at the back-end of application server. 
The pathways corresponding to petty thief are shown in Fig. \ref{petty-thief-pathways}. It is to be noted here that some of the actions mentioned in the figure can not be captured by our testbed itself and have to be enabled separately.

\subsubsection{Black-hat pathways}
\begin{figure}[h]
\centering
\includegraphics[scale=0.25]{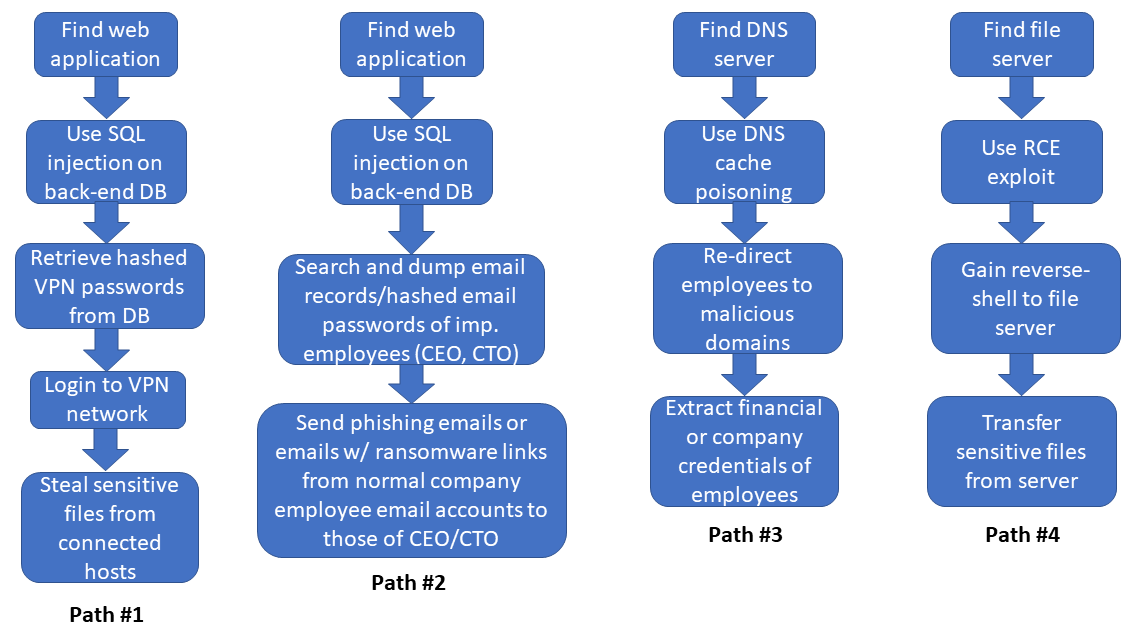}
\caption{Possible testbed pathways for a Black-hat}
\label{prof-pathways}
\end{figure}

Black-hats are highly skilled and are typically after or hired to steal high-worth information. Targeting large enterprise networks, similar to hacktivists, they may have to use malware to spread and infect machines, use privilege escalation techniques and move across the company network (LAN2 $\rightarrow$ LAN1). Once they have established access to both the LANs, they may attempt to steal confidential high-value product files (e.g., source code) stored on hosts connected to the company's remote-site VPN network, or send phishing emails with ransomware/malware attachments from legitimate employee e-mail accounts to the e-mail address of a high-value target such as the company CEO/CTO (to make them look convincing) and extract valuable information once they get access to the target's computer. The pathways corresponding to black-hat are shown in Fig. \ref{prof-pathways}.


\section{Preliminary Data Collection and Analysis}
\label{data-analysis}
We implemented our proposed testbed on a VMware ESXi server. The same testbed can also be configured on a cyber security experimentation platform such as \textit{DETERlab}. The testbed machines are Virtual Machines (VMs) configured on the ESXi server and running either Ubuntu 16.04, Ubuntu 20.04 or Windows 10 OS. Most VMs are configured with \textasciitilde 8GB of RAM, 4 virtual CPUs and few hundreds of GBs of hard disk. The web servers are built using \textit{Apache2}, database using \textit{MySQL}, DNS server using \textit{BIND9} and file server using \textit{Samba}. All the VMs except the VPN host have their system times synchronized to a local NTP (Network Time Protocol) server. The VPN server broadcasts the time on the VPN network which is received by VPN host. All the VMs have an administrator account (representing the company's IT admin account) and few VMs have a local user account (representing the employee's account). A virtual NAT router provides access to the Internet for all the testbed VMs. One of the VMs is also configured with an \textit{OPNsense} firewall to manage Internet access for the other VMs. By default, external Internet access is blocked for all the VMs except the jump host.

\begin{figure}[h]
\centering
\includegraphics[scale=0.25]{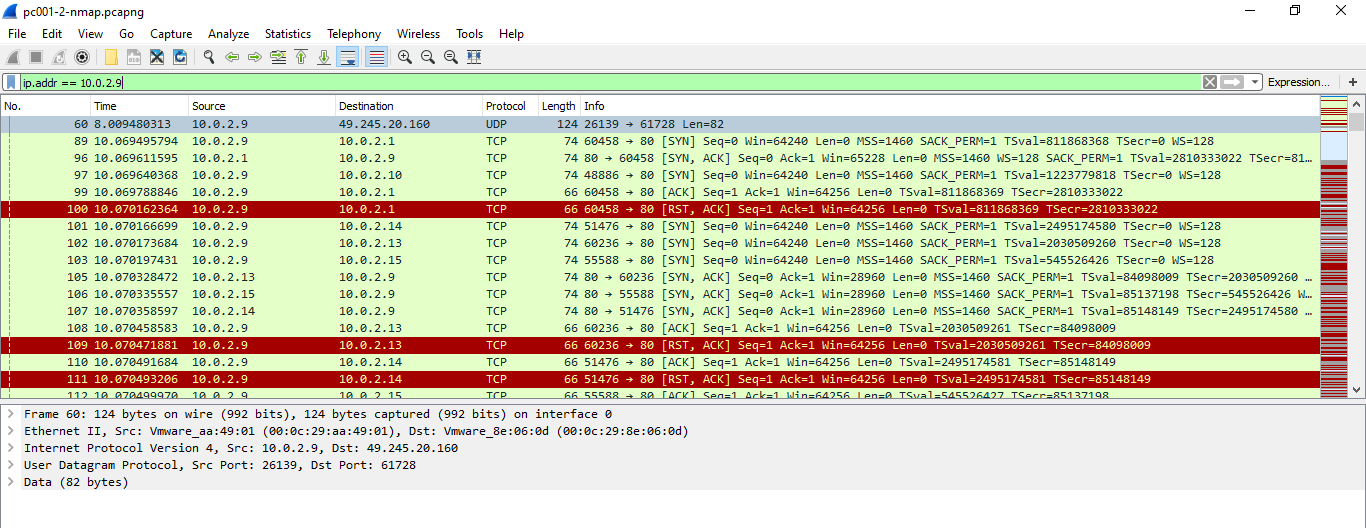}
\caption{Packet capture at logging server during \textit{nmap} scan}
\label{nmap-pcap}
\end{figure}

One of the first actions by any hacker who targets a network, enterprise or not, is to find other hosts connected to the same subnet. We therefore run an open-source network scanning tool, \textit{nmap} on the jump host using the subnet (\textit{10.0.2.0/24}). It is reflected in the packet capture collected on logging server as shown in Fig. \ref{nmap-pcap}. The jump host makes TCP connection requests to all the connected hosts in the subnet, followed by a three-way handshake ([SYN], [SYN,ACK], [ACK]) for each established connection. It also makes DNS queries to the Google DNS server (IP address: \textit{8.8.8.8}) to obtain domain name mappings for the connected hosts if any. Thus, if in an enterprise network, there is an abnormal increase in number of TCP connection requests to hosts or number of DNS queries, it is likely due to someone trying to scan the network.

\begin{figure}[h]
\centering
\includegraphics[scale=0.25]{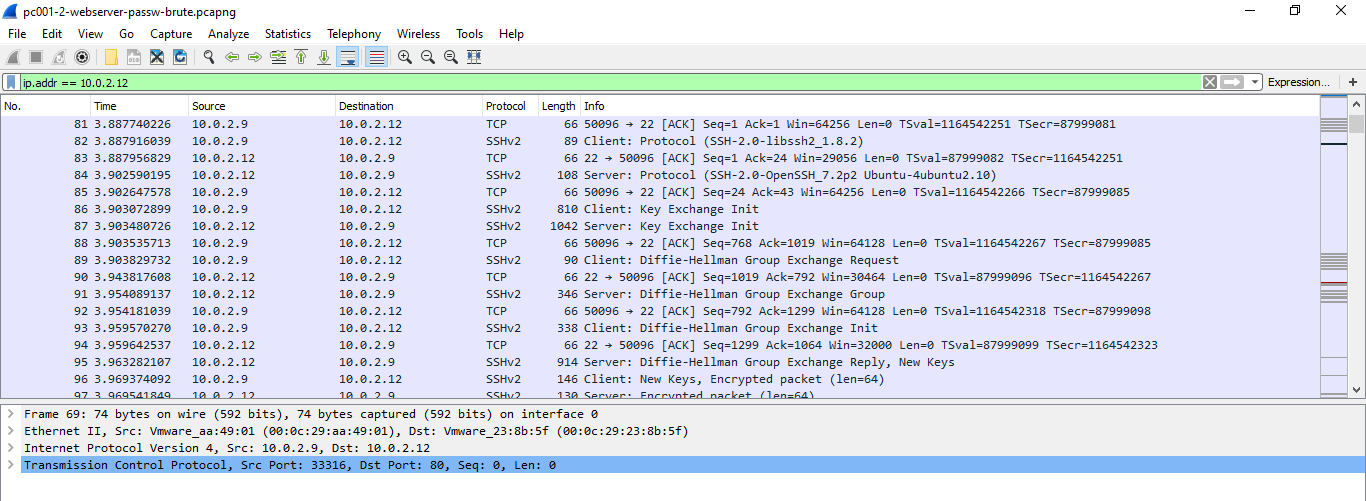}
\caption{Packet capture at logging server during during credential brute forcing on web server}
\label{webserver-passw-brute}
\end{figure}

If a hacker find the web server, he/she may attempt to brute-force its credentials to login to the server via SSH using tools such as \textit{metasploit}/\textit{hydra}/\textit{nmap}-scripting engine. Login credential brute-forcing is a common tactic employed by hackers when they suspect a vulnerable web server. As seen from the packet capture at logging server displayed in Fig. \ref{webserver-passw-brute}, such as action would result in multiple TCP connection requests from the jump host to the target web server, followed by a three-way handshake ([SYN], [SYN,ACK], [ACK] packets) for each established connection. It also leads to client-server Diffie-Hellman key exchanges followed by forwarding of encrypted packets containing \textit{username} and \textit{password} from client to server. As most of the password brute-force attempts are unsuccessful and SSH allows a maximum of 6 authentication attempts per connection, it leads to several connection resets ([RST] packets). This means that if there is an abnormal increase in number of TCP connection requests to a host (particularly a web server), DH key exchanges and TCP connection resets, it is likely due to someone brute-forcing SSH login credentials for the host.

\begin{figure}[h]
\centering
\includegraphics[scale=0.25]{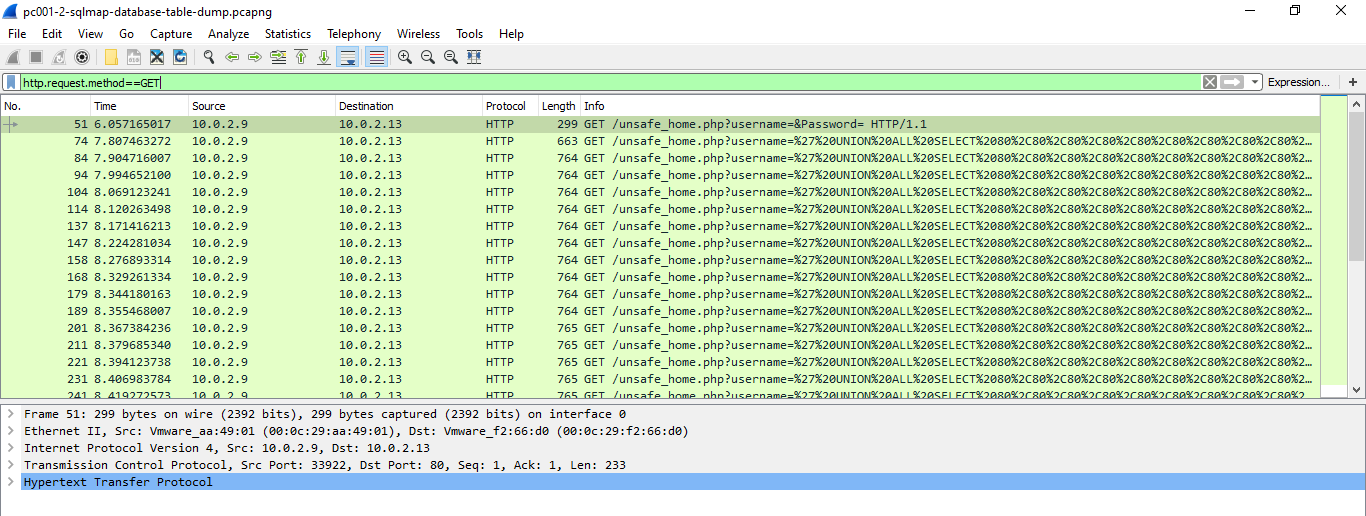}
\caption{Packet capture at logging server during SQL injection attack on web application database}
\label{sqlmap-database-dump}
\end{figure}

A hacker may also find the web application server with database at its back-end by probing the testbed network. Since SQL injection is one of the most common attacks carried out by hackers against databases, we run an open-source penetration testing tool for automated SQL injection vulnerability detection and exploitation, \textit{sqlmap} on the jump host to find if the web application database is vulnerable to SQL injection and subsequently, dump the complete database. A packet capture during the attack (Fig. \ref{sqlmap-database-dump}) shows multiple TCP connection requests followed by \textit{HTTP GET} requests sent from the jump host to the web application server. It can thus be inferred that an anomalous increase in the number of TCP connection requests and HTTP GET requests to a host (particularly a web application) is indicative of someone dumping the back-end database.

\section{Conclusion} 
We have proposed a testbed to analyse the attack strategies of cyber adversaries in enterprise networks and use them for attack detection. The testbed network consists of several nodes with different functionalities, with some of the nodes infected with security vulnerabilities. The testbed presents multiple pathways to a hacker invited to attack the testbed. We discuss an implementation of our proposed testbed, subject it to few well-known cyber-attack strategies and analyse the data collected. Our preliminary analysis strengthens the initial argument presented earlier that data collected from the testbed can be used to capture patterns in attack strategies deployed by adversaries in enterprise networks.



\bibliographystyle{ACM-Reference-Format}
\bibliography{ctftestbed}


\end{document}